# STOCK EXCHANGE SHARES RANKING AND BINARY-TERNARY COMPRESSIVE CODING

# РАНЖИРОВАНИЕ ИНВЕСТИЦИОННОЙ ПРИВЛЕКАТЕЛЬНОСТИ АКЦИЙ И ДВОИЧНО-ТРОИЧНОЕ СЖИМАЮЩЕЕ КОДИРОВАНИЕ

**Igor Nesiolovskiy**
*Podolsk, Moscow Region, Russia*
nesiolovskiy@gmail.com

**Игорь Несиоловский**
*Россия, Московская область, Подольск*
nesio@mail.ru

## Abstract (Аннотация)

This paper proposes a method for ranking the investment attractiveness of exchange-traded stocks where investment risk is not related to the volatility indicator but instead is related to the indicator of compression of the time series of price changes. The article describes in detail the ranking algorithm, provides an example of ranking the shares of all companies included in the Dow Jones stock index. The paper additionally compares the results of ranking these stocks by volatility and compression and also shows the strengths of the second indicator, which is formed using the method of binary-ternary compression of historical financial data.

В данной работе предлагается метод ранжирования инвестиционной привлекательности ценных бумаг фондового рынка, в котором их инвестиционный риск оценивается не по традиционному показателю волатильности доходности, а по показателю сжатия временных рядов данных изменения их биржевой цены. В работе подробно описан алгоритм ранжирования, представлен пример ранжирования акций всех компаний, входящих в фондовый индекс Dow Jones, дополнительно проведено сравнение результатов ранжирования этих акций по показателям волатильности и сжатия и отмечены сильные стороны второго показателя, формируемого с использованием метода двоично-троичного сжимающего кодирования исторических данных биржевых котировок ценных бумаг.

***Ключевые слова:*** *ценные бумаги биржевого фондового рынка, ранжирование, анализ временных рядов, сжатие, двоично-троичное кодирование данных.*

# 1 Введение (Introduction)

При принятии решений об инвестировании в ценные бумаги фондового рынка используют различные инструменты технического, фундаментального, портфельного анализа. Технический анализ основывается на изучении регулярных фактических данных, формируемых фондовым рынком – прежде всего исторических (или временны́х) рядов данных котировок и объёмов торгов. К основным задачам такого изучения относят выявление существующих закономерностей и прогнозирование будущих значений исследуемых рядов. Для прогнозирования предлагаются различные методы из областей математической статистики, машинного обучения, универсального кодирования или «сжатия данных». Последний класс методов представлен, например, работами [1], [2] разных лет. В настоящей работе предлагается метод этого же класса, но применительно к анализу закономерностей исторических данных фондового рынка ценных бумаг.

# 2 Концептуальные положения (Conceptual provisions)

На одном из этапов технического анализа целесообразно ранжировать ценные бумаги, представленные на рынке, по их инвестиционной привлекательности. Для этого следует оценить и сопоставить между собой прежде всего такие финансовые показатели потенциальных объектов инвестирования как *рискованность*, *ликвидность*, *доходность*.

Оценка *риска*, связанного с приобретением возможного инвестиционного актива, реализуется с использованием метода двоично-троичного сжимающего кодирования данных временных рядов изменений биржевых котировок соответствующей ценной бумаги.

Формализм и примеры использования указанного метода кодирования вне связи с финансовой математикой подробно описаны в [3] и [4]. В общем случае он является математическим обеспечением автоматизированных процессов сжимающего кодирования исходных данных без потерь, позволяющим уменьшить информационную избыточность в следующих друг за другом серийных элементах («буквах») исходных данных (текстового, числового и других видов представления информации). Уменьшение избыточности достигается за счёт применения специальных наборов сжимающих двоично-троичных кодов, заменяющих чаще встречающиеся серийные элементы исходных данных более короткими кодами, и наоборот, реже встречающие элементы – более длинными кодами. Результат кодирования данных по этому методу характеризуется меньшими объёмами их хранения или передачи, чем требуется для исходных данных. Метод определяет также и технику декодирования данных, предварительно закодированных в соответствии с ним же.

Основная идея его применения к биржевым котировкам ценных бумаг заключается в том, что исходные числовые данные изменений цены предварительно фильтруются (сглаживаются) и преобразуются к символьному виду, а затем кодируются символами определённого символьного алфавита, приводящими к сжатию исходных данных. Чем выше показатель сжатия, тем больше закономерностей, содержащихся в исходных данных, учтено и использовано алгоритмом метода при их обработке. Если эффективность сжатия временного ряда изменений котировок одной ценной бумаги превосходит эффективность сжатия подобного временного ряда другой ценной бумаги, то справедливо считать, что в рассматриваемом периоде анализа поведение биржевого курса первой бумаги было более закономерным, а значит менее случайным, чем второй. Поэтому инвестиционные вложения в неё скорее всего будут менее рискованными, и такой вывод будет достаточно обоснованным, если анализ временных рядов исходных данных проводился за довольно длительный исторический отрезок времени функционирования фондового рынка.

*Ликвидность* ценной бумаги может оцениваться по историческим данным объёмов торгов, где высокий спрос и предложение на неё свидетельствует о её высокой востребованности на рынке. Положительная *доходность* бумаги может характеризоваться наличием восходящего тренда её биржевой цены, формируемого результатами состоявшихся в прошлом биржевых торгов.

Рассматриваемый ниже метод ранжирования инвестиционной привлекательности ценных бумаг ориентирован на поддержку долгосрочной (и в некоторой степени консервативной) стратегии вложения финансовых средств в рынок акций или ценных бумаг биржевых фондов.

## 3   Алгоритм ранжирования (Ranking algorithm)

Рассмотрим алгоритм метода ранжирования в виде, наиболее удобном для его последующей программной реализации (разработки).

**1.** Пусть имеется временной ряд $P$ котировок (price) ценной бумаги $A$ (например, на момент закрытия ежедневных торгов) $P_A\{p_1, p_1, p_1, ..., p_n\}$, где $p_i$ характерная котировка $i$-того торгового дня. Значение $n$ устанавливается достаточно большим – например, соответствует 5 (пяти) последним годам проведения торгов (относительно календарной даты выполнения ранжирования). Среди членов указанного ряда существует один или несколько элементов с максимальным значением цены $p_{max}$.

**2.** Сформируем модифицированный ряд данных с приведённой (или относительной) ценой и элементами

$$\bar{p}_i = [\,p_i/p_{max} * 100\,],$$

значения которых округлены до целого, где $i$ изменяется от 1 до $n$.

**3.** Образуем ряд изменений относительной цены с элементами

$$\bar{\bar{p}}_i = \bar{p}_{i+1} - \bar{p}_i, \quad \text{где } i \text{ изменяется от } 1 \text{ до } n-1.$$

**4.** Преобразуем полученный ряд числовых значений в символьную форму, подлежащую в дальнейшем двоично-троичному сжимающему кодированию. Для этого воспользуемся следующим трёхсимвольным шаблоном преобразования каждого члена ряда изменений относительной цены:

$$\bar{\bar{p}}_k \Rightarrow XYZ, \quad \text{где}$$

- $X$ – обязательный подстановочный символ, принимающий значение символа «−» (минус), если $\bar{\bar{p}}_k < 0$; значение символа «~» (тильда), если $\bar{\bar{p}}_k = 0$; значение символа «+» (плюс), если $\bar{\bar{p}}_k > 0$;
- $YZ$ – обязательная подстановочная пара символов, представляющих два первых числовых разряда значения $\bar{\bar{p}}_k$ в форме с так называемым ведущим нулём, который применяется в случае, если $|\bar{\bar{p}}_k| < 10$.

Например, $\bar{\bar{p}}_k = -11 \Rightarrow$ «–11»; $\bar{\bar{p}}_k = -3 \Rightarrow$ «–03»; $\bar{\bar{p}}_k = 0 \Rightarrow$ «~00»; $\bar{\bar{p}}_k = 1 \Rightarrow$ «+01»; $\bar{\bar{p}}_k = 10 \Rightarrow$ «+10».

**5.** В результате символьного преобразования ряда мы получили его символьное представление $S$ (например, с фрагментом –11–03~00+01+10). Приступаем к выполнению его двоично-троичного сжимающего кодирования. В качестве входного параметра алгоритма кодирования определяем, что алфавит исходных данных состоит из трёхсимвольных серийных элементов вида $XYZ$, рассмотренных выше. Поскольку схема кодирования подробно описана в указанных выше первоисточниках, то здесь она не рассматривается. При выполнении её алгоритма нас будут интересовать (в контексте описываемой в данной работе задачи ранжирования) три показателя:

- $l_{src}$ – исходная (source) длина символьного представления $S$ временного ряда $P$ ценной бумаги $A$, *бит*;
- $l_{cod}$ – результирующая длина закодированного (code) двоично-троичного представления символьного ряда $S$, *бит*;
- $l_{abc}$ – общая длина символьного алфавита (abc) исходных данных, уникальные элементы (буквы) которого представлены согласно трёхсимвольному шаблону $XYZ$, *бит*.

**6.** Для каждой из $m$ ценных бумаг, отобранных в листинг (список) ранжирования, вычисляем целевой показатель сжатия $r_A$:

$$r_A = \left( \left( \frac{l_{cod} + l_{abc}}{l_{src}} \right)_A * 100 \right) * \frac{n_{Max}}{n_A}, \quad \text{где}$$

- $A$ изменяется от 1 до $m$ (здесь $m$ количество акций в списке ранжирования);

- $\left(\frac{l_{cod}+l_{abc}}{l_{src}}\right)_A$ – степень сжатия символьного представления $S$ временного ряда $P$ ценной бумаги $A$;
- $n_A$ – количество членов временного ряда $P$ ценной бумаги $A$;
- $n_{Max}$ – количество членов максимально длинного временного ряда у некоторой ценной бумаги из списка рядов $P_1, P_2, P_3, \ldots, P_m$, которое требуется предварительно определить, где временные ряды всех бумаг списка построены по данным ежедневных торгов за предыдущие пять лет (тем самым допускается, что некоторые бумаги из списка ранжирования могут иметь более короткую историю торгов на фондовом рынке, чем остальные).

В формуле для $r_A$ допустима замена отношения $\frac{n_{Max}}{n_A}$ на отношение $\frac{n_{Max}-1}{n_A-1}$.

**7.** Каждое из полученных значений $r_A$ умножаем на знакопеременную функцию $sgn$, которая принимает значения:

- $sgn = -1$, если для исходного временного ряда данных $P$ ценной бумаги $A$ выполняется хотя бы одно из неравенств: $p_1 > p_n$ или $p_1 > p_{avg}$ или $p_{avg} > p_n$, где $p_{avg}$ – среднее арифметическое всех значений временного ряда;
- $sgn = 1$, если не выполняется ни одно из указанных выше неравенств для $p_1, p_{avg}, p_n$.

**8.** Сортируем подмножество всех значений $r_A$, для которых справедливо условие $r_A > 0$, по возрастанию и присваиваем им порядковые номера, начиная с 1. Если у двух ценных бумаг $A$ и $B$ величины показателей сжатия равны ($r_A = r_B$), то они нумеруются одинаковым значением. Такое правило нумерации справедливо и для нескольких бумаг, если их показатели сжатия равны между собой (при этом здесь и далее обеспечивается условие, что значение ранга ценной бумаги в листинге ранжирования будет всегда целочисленным, которое, вообще говоря, с методической точки зрения не является единственно допустимым).

**9.** Сортируем подмножество всех значений $r_A$, для которых справедливо условие $r_A < 0$, по убыванию и присваиваем им порядковые номера, продолжая нумерацию, выполненную на предыдущем шаге алгоритма, и сохраняя логику такой нумерации.

На этом ранжирование списка ценных бумаг по показателю сжатия завершено. Переходим к ранжированию ценных бумаг по показателю денежного потока.

**10.** Пусть имеется временной ряд $V$ ежедневных объёмов торгов (volume) ценной бумаги $A$ $V_A\{v_1, v_1, v_1, \ldots, v_n\}$, синхронный по датам торгов с рассмотренным выше рядом котировок $P_A\{p_1, p_1, p_1, \ldots, p_n\}$, где $v_i$ объём торгов $i$-того торгового дня.

**11.** Образуем ряд $C$ значений условного денежного (cash flow) с элементами $c_i = v_i * p_i$, где $i$ изменяется от 1 до $n$.

**12.** Для каждой из $m$ ценных бумаг, отобранных в листинг (список) ранжирования, вычисляем среднее квадратическое значение условного денежного потока $\bar{c}_A$:

$$\bar{c}_A = \sqrt{\frac{1}{n_A} * (c_1^2 + \cdots + c_n^2)_A}, \text{ где}$$

- $A$ изменяется от 1 до $m$ (здесь $m$ количество ценных бумаг в списке ранжирования);
- $n_A$ – количество членов временного ряда $C$ ценной бумаги $A$.

**13.** Сортируем множество всех значений $\bar{c}_A$ по убыванию и присваиваем им порядковые номера, начиная с 1.

На этом ранжирование списка ценных бумаг по показателю условного денежного потока завершено. Переходим к итоговому ранжированию ценных бумаг.

**14.** Для каждой из ценных бумаг из списка ранжирования суммируем значения ранга по показателю сжатия (см. шаги 8-9 рассматриваемого алгоритма) и ранга по показателю денежного потока (см. шаг 13), после чего сортируем список бумаг по возрастанию значения полученного суммарного ранга.

На этом формирование листинга ранжирования ценных бумаг фондового рынка завершено.

## 4 Пример расчёта показателей ранжирования (Example of calculating ranking indicators)

Для иллюстрации использования описанного выше метода выполнено ранжирование акций, входящих в фондовый индекс Dow Jones, который охватывает 30 крупнейших компаний США. Результаты ранжирования представлены в следующей ниже таблице.

При ранжировании в качестве исходных были использованы исторические ряды данных о ежедневных ценах закрытия (price) и объёмах продаж ценных бумаг этих компаний за период с 14 октября 2016 г. по 13 октября 2021 г. (5-летний период), предоставляемые web-сервисом Google Finance.

В процессе обработки данных использовалась функциональность электронных таблиц Google Sheets и Microsoft Excel и средства программирования Google Apps Scripts и Visual Basic for Application.

В списке ранжирования только одна бумага (стикер DOW) выделяется более короткой историей торгов (согласно данным Google Finance) по сравнению с остальными.

# Stock Exchange Shares Ranking And Binary-Ternary Compressive Coding

**ЛИСТИНГ РАНЖИРОВАНИЯ ЦЕННЫХ БУМАГ, ВХОДЯЩИХ В ИНДЕКС DOW JONES**

| 14-10-2016 | 13-10-2021 | - период анализа |
| | DAILY | - детализация исходных данных |
| | USD | - валюта анализа |

| Название индекса и код ценной бумаги (sticker) | Название компании (company) | Длина ряда исходных данных Цены закрытия (points) | Показатель сжатия ряда данных изменения Цены закрытия, % (variaty) | Микро-график ряда исходных данных Цены закрытия (microchart) | Рост Цены закрытия за период, % (growth) | Ранг аналитики по Цене закрытия (rank) | Ранг аналитик суммарный (total rank) | Ранг аналитики по Денежному потоку (rank) | Длина ряда исходных данных Объема торгов (points) | Среднее квадратическое Денежного потока (root-mean-square) |
|---|---|---|---|---|---|---|---|---|---|---|
| DOWJONES_MSFT | Microsoft Corporation | 1257 | 16,00318471 | | 410 | 1 | 3 | 2 | 1257 | 4 924 657 291 |
| DOWJONES_AAPL | Apple Inc | 1257 | 16,17569002 | | 381 | 2 | 3 | 1 | 1257 | 6 931 222 546 |
| DOWJONES_V | Visa Inc | 1257 | 16,54723992 | | 172 | 8 | 13 | 5 | 1257 | 1 471 517 711 |
| DOWJONES_CRM | salesforcecom inc | 1257 | 16,40127389 | | 276 | 6 | 15 | 9 | 1257 | 1 338 206 283 |
| DOWJONES_HD | Home Depot Inc | 1257 | 16,4410828 | | 167 | 7 | 18 | 11 | 1257 | 1 040 156 295 |
| DOWJONES_JPM | JPMorgan Chase & Co | 1257 | 16,86571125 | | 145 | 14 | 18 | 4 | 1257 | 1 784 428 048 |
| DOWJONES_DIS | Walt Disney Co | 1257 | 16,85244161 | | 90 | 13 | 20 | 7 | 1257 | 1 559 375 992 |
| DOWJONES_GS | Goldman Sachs Group Inc | 1257 | 16,30838641 | | 127 | 4 | 21 | 17 | 1257 | 822 365 723 |
| DOWJONES_UNH | UnitedHealth Group Inc | 1257 | 16,62685775 | | 201 | 9 | 21 | 12 | 1257 | 1 070 129 583 |
| DOWJONES_WMT | Walmart Inc | 1257 | 16,79936306 | | 104 | 11 | 24 | 13 | 1257 | 1 032 778 856 |
| DOWJONES_CAT | Caterpillar Inc | 1257 | 16,28184713 | | 117 | 3 | 26 | 23 | 1257 | 677 779 812 |
| DOWJONES_NKE | Nike Inc | 1257 | 16,33492569 | | 197 | 5 | 26 | 21 | 1257 | 759 956 485 |
| DOWJONES_BA | Boeing Co | 1257 | {16,53397028} | | 67 | 24 | 27 | 3 | 1257 | 3 339 994 965 |
| DOWJONES_CSCO | Cisco Systems Inc | 1257 | 21,32430998 | | 80 | 17 | 27 | 10 | 1257 | 1 056 056 795 |

STOCK EXCHANGE SHARES RANKING AND BINARY-TERNARY COMPRESSIVE CODING

| Название индекса и код ценной бумаги (sticker) | Название компании (company) | Длина ряда исходных данных Цены закрытия (points) | Показатель сжатия ряда данных изменения Цены закрытия, % (variaty) | Микро-график ряда исходных данных Цены закрытия (microchart) | Рост Цены закрытия за период, % (growth) | Ранг аналитики по Цене закрытия (rank) | Ранг аналитик суммарный (total rank) | Ранг аналитики по Денежному потоку (rank) | Длина ряда исходных данных Объема торгов (points) | Среднее квадратическое Денежного потока (root-mean-square) |
|---|---|---|---|---|---|---|---|---|---|---|
| DOWJONES_INTC | Intel Corporation | 1257 | 21,70912951 | | 39 | 21 | 27 | 6 | 1257 | 1 497 920 089 |
| DOWJONES_JNJ | Johnson & Johnson | 1257 | 21,47027601 | | 34 | 20 | 28 | 8 | 1257 | 1 101 848 089 |
| DOWJONES_PG | Procter & Gamble Co | 1257 | 16,985138 | | 61 | 15 | 29 | 14 | 1257 | 937 299 533 |
| DOWJONES_MCD | McDonald s Corp | 1257 | 16,73301486 | | 114 | 10 | 32 | 22 | 1257 | 700 537 424 |
| DOWJONES_MRK | Merck & Co Inc | 1257 | 21,43046709 | | 34 | 19 | 37 | 18 | 1257 | 843 885 281 |
| DOWJONES_AXP | American Express Company | 1257 | 16,83917197 | | 191 | 12 | 39 | 27 | 1257 | 460 478 802 |
| DOWJONES_HON | Honeywell International Inc | 1257 | 16,86571125 | | 105 | 14 | 40 | 26 | 1257 | 536 096 677 |
| DOWJONES_KO | Coca-Cola Co | 1257 | 21,72239915 | | 30 | 22 | 41 | 19 | 1257 | 736 051 379 |
| DOWJONES_AMGN | Amgen Inc | 1257 | 21,33757962 | | 25 | 18 | 42 | 24 | 1257 | 679 059 594 |
| DOWJONES_CVX | Chevron Corporation | 1257 | {21,31104034} | | 6 | 27 | 43 | 16 | 1257 | 900 795 672 |
| DOWJONES_VZ | Verizon Communications Inc | 1257 | {21,80201699} | | 2 | 29 | 44 | 15 | 1257 | 910 789 686 |
| DOWJONES_TRV | Travelers Companies Inc | 1257 | 17,13110403 | | 36 | 16 | 46 | 30 | 1257 | 211 646 001 |
| DOWJONES_IBM | IBM Common Stock | 1257 | {21,32430998} | | -9 | 28 | 48 | 20 | 1257 | 774 301 777 |
| DOWJONES_MMM | 3M Co | 1257 | {16,87898089} | | 3 | 26 | 51 | 25 | 1257 | 529 107 096 |
| DOWJONES_DOW | Dow Inc | 648 | 36,07795714 | | 16 | 23 | 52 | 29 | 648 | 272 400 059 |
| DOWJONES_WBA | Walgreens Boots Alliance Inc | 1257 | {16,65339703} | | -39 | 25 | 53 | 28 | 1257 | 414 246 923 |

Серой заливкой и фигурными скобками в таблице отмечены отрицательные значения показателя сжатия, которые формируются алгоритмом. Такие бумаги в целом отличаются малопривлекательным (а некоторые даже негативным) трендом 5-летней доходности, но они ранжировались в списке на общих основаниях.

Рассмотренный метод ранжирования соотносит инвестиционный риск как меру предсказуемости/непредсказуемости (закономерности/стохастичности) поведения цены, который выражен показателем сжатия временного ряда изменений относительной цены закрытия торгов ценной бумаги (или ряда значений доходности), и ликвидность как меру спроса на ценную бумагу и возможности её быстрой реализации на фондовом рынке, которая выражена показателем условного денежного потока. В представленном примере в суммарном ранге этих двух аналитических показателей веса их индивидуальных рангов приняты одинаковыми.

В листинге ранжирования можно наблюдать, что по мере убывания ранга инвестиционной привлекательности ценных бумаг содержащиеся в их микрографиках положительные (восходящие) тренды котировок постепенно сменяются на нейтральные (более пологие) и даже на отрицательные (нисходящие). Так же по графикам можно судить, что «искривлённость» («изломанность») линий поведения цены акций увеличивается по мере увеличения их суммарного ранга, но эту тенденцию стремятся нарушить значения ранга показателя денежного потока отдельных акций. По листингу заметно, что увеличение суммарного ранга в целом коррелирует со снижением доходности и ликвидности ценных бумаг.

## 5 Сравнение ранжирования по показателям сжатия и волатильности (Comparison of the ranking by compression and volatility indicators)

Общепринято, что показатель *волатильности* ценной бумаги используется как мера риска инвестиционных вложений в неё. Поскольку в рассматриваемой в данной работе методике предлагается использовать альтернативную меру риска, то полезно сравнить их между собой – показатели волатильности цены и сжатия ряда данных изменения цены инвестиционного инструмента, для расчёта которых используются одни и те же исходные данные.

Рассчитаем волатильность одним из общеизвестных способов.

**1.** Для каждой ценной бумаги сформируем модифицированный ряд данных с приведённой (или относительной) ежедневной доходностью

$$\hat{p}_i = p_{i+1}/p_i - 1,$$

где $i$ изменяется от $1$ до $n-1$ (здесь $n$ количество значений временного ряда $P$ ценной бумаги $A$.

**РАНЖИРОВАНИЕ ЦЕННЫХ БУМАГ, ВХОДЯЩИХ В ИНДЕКС DOW JONES, ПО ПОКАЗАТЕЛЮ ВОЛАТИЛЬНОСТИ ДОХОДНОСТИ**

| 14-10-2016 | 13-10-2021 | - период анализа |
| | DAILY | - детализация исходных данных |
| | USD | - валюта анализа |

| Название индекса и код ценной бумаги (sticker) | Название компании (company) | Длина ряда исходных данных Цены закрытия (points) | Волатильность доходности по данным ряда Цены закрытия, % (volatility) | Микро-график ряда исходных данных Цены закрытия (microchart) | Рост Цены закрытия за период, % (growth) | Ранг аналитики (rank) |
|---|---|---|---|---|---|---|
| DOWJONES_VZ | Verizon Communications Inc | 1257 | 42,53636 | | 2 | 1 |
| DOWJONES_JNJ | Johnson & Johnson | 1257 | 44,79160 | | 34 | 2 |
| DOWJONES_KO | Coca-Cola Co | 1257 | 44,92054 | | 30 | 3 |
| DOWJONES_PG | Procter & Gamble Co | 1257 | 45,32519 | | 61 | 4 |
| DOWJONES_WMT | Walmart Inc | 1257 | 48,61491 | | 104 | 5 |
| DOWJONES_MRK | Merck & Co Inc | 1257 | 49,23611 | | 34 | 6 |
| DOWJONES_MCD | McDonald s Corp | 1257 | 51,56944 | | 114 | 7 |
| DOWJONES_MMM | 3M Co | 1257 | 55,66549 | | 3 | 8 |
| DOWJONES_HON | Honeywell International Inc | 1257 | 56,14899 | | 105 | 9 |
| DOWJONES_AMGN | Amgen Inc | 1257 | 57,12106 | | 25 | 10 |
| DOWJONES_HD | Home Depot Inc | 1257 | 57,35623 | | 167 | 11 |
| DOWJONES_V | Visa Inc | 1257 | 58,06310 | | 172 | 12 |
| DOWJONES_IBM | IBM Common Stock | 1257 | 58,33645 | | -9 | 13 |
| DOWJONES_CSCO | Cisco Systems Inc | 1257 | 60,03519 | | 80 | 14 |
| DOWJONES_TRV | Travelers Companies Inc | 1257 | 60,26622 | | 36 | 15 |
| DOWJONES_MSFT | Microsoft Corporation | 1257 | 61,18618 | | 410 | 16 |
| DOWJONES_UNH | UnitedHealth Group Inc | 1257 | 63,76063 | | 201 | 17 |
| DOWJONES_DIS | Walt Disney Co | 1257 | 64,02403 | | 90 | 18 |
| DOWJONES_NKE | Nike Inc | 1257 | 64,87397 | | 197 | 19 |
| DOWJONES_JPM | JPMorgan Chase & Co | 1257 | 67,31923 | | 145 | 20 |
| DOWJONES_AAPL | Apple Inc | 1257 | 67,53565 | | 381 | 21 |

| Название индекса и код ценной бумаги (sticker) | Название компании (company) | Длина ряда исходных данных Цены закрытия (points) | Волатильность доходности по данным ряда Цены закрытия, % (volatility) | Микро-график ряда исходных данных Цены закрытия (microchart) | Рост Цены закрытия за период, % (growth) | Ранг аналитики (rank) |
|---|---|---|---|---|---|---|
| DOWJONES_WBA | Walgreens Boots Alliance Inc | 1257 | 69,65455 | | -39 | 22 |
| DOWJONES_CAT | Caterpillar Inc | 1257 | 69,67088 | | 117 | 23 |
| DOWJONES_GS | Goldman Sachs Group Inc | 1257 | 70,61306 | | 127 | 24 |
| DOWJONES_CVX | Chevron Corporation | 1257 | 73,81625 | | 6 | 25 |
| DOWJONES_CRM | salesforcecom inc | 1257 | 74,15989 | | 276 | 26 |
| DOWJONES_AXP | American Express Company | 1257 | 76,33491 | | 191 | 27 |
| DOWJONES_INTC | Intel Corporation | 1257 | 76,45159 | | 39 | 28 |
| DOWJONES_DOW | Dow Inc | 648 | 102,78006 | | 16 | 29 |
| DOWJONES_BA | Boeing Co | 1257 | 103,98828 | | 67 | 30 |

**2.** Для каждой из $m$ ценных бумаг, отобранных в листинг ранжирования, рассчитаем среднее арифметическое ежедневной доходности $\bar{p}_A$:

$$\bar{p}_A = ( (\hat{p}_1 + \cdots + \hat{p}_{n-1})/(n-1) )_A , \quad \text{где}$$

- $A$ изменяется от 1 до $m$ (здесь $m$ количество ценных бумаг в списке ранжирования);
- $n$ количество значений временного ряда $P$ ценной бумаги $A$.

**3.** Для каждой из $m$ ценных бумаг определяем среднее квадратическое (стандартное) отклонение ежедневной доходности $\bar{\delta}_A$:

$$\bar{\delta}_A = \sqrt{( (\hat{p}_1 - \bar{p}_A)^2 + \cdots + (\hat{p}_{n-1} - \bar{p}_A)^2 )/(n-1) )_A} \quad, \text{где } A \text{ и } n \text{ определены выше.}$$

**4.** Для каждой ценной бумаги $A$ рассчитываем волатильность $\delta_A$ за весь период анализа:

$$\delta_A = \bar{\delta}_A * \sqrt{n_{Max} - 1} \quad, \text{где}$$

- $n_{Max}$ – количество членов максимально длинного временного ряда у некоторой ценной бумаги из списка рядов $P_1, P_2, P_3, \ldots, P_m$, которую требуется предварительно определить, где временные ряды всех бумаг списка построены по данным ежедневных торгов за предыдущие пять лет (тем самым допускается, что некоторые бумаги из списка ранжирования могут иметь более короткую историю торгов на фондовом рынке, чем остальные).

**5.** Сортируем множество всех значений $\delta_A$ по возрастанию и присваиваем им порядковые номера, начиная с 1.

РАНЖИРОВАНИЕ ЦЕННЫХ БУМАГ, ВХОДЯЩИХ В ИНДЕКС DOW JONES, ПО ПОКАЗАТЕЛЮ СЖАТИЯ ИЗМЕНЕНИЙ ЦЕНЫ ЗАКРЫТИЯ

| 14-10-2016 | 13-10-2021 | - период анализа | | | | |
| | DAILY | - детализация исходных данных | | | | |
| | USD | - валюта анализа | | | | |

| Название индекса и код ценной бумаги (sticker) | Название компании (company) | Длина ряда исходных данных Цены закрытия (points) | Показатель сжатия ряда данных изменений Цены закрытия, % (variaty) | Микро-график ряда исходных данных Цены закрытия (microchart) | Рост Цены закрытия за период, % (growth) | Ранг аналитики (rank) |
|---|---|---|---|---|---|---|
| DOWJONES_MSFT | Microsoft Corporation | 1257 | 16,00318 | 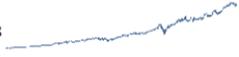 | 410 | 1 |
| DOWJONES_AAPL | Apple Inc | 1257 | 16,17569 | 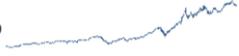 | 381 | 2 |
| DOWJONES_CAT | Caterpillar Inc | 1257 | 16,28185 | 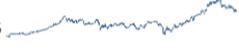 | 117 | 3 |
| DOWJONES_GS | Goldman Sachs Group Inc | 1257 | 16,30839 | 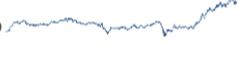 | 127 | 4 |
| DOWJONES_NKE | Nike Inc | 1257 | 16,33493 | 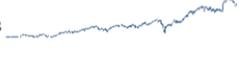 | 197 | 5 |
| DOWJONES_CRM | salesforcecom inc | 1257 | 16,40127 | 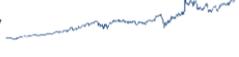 | 276 | 6 |
| DOWJONES_HD | Home Depot Inc | 1257 | 16,44108 | 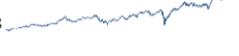 | 167 | 7 |
| DOWJONES_V | Visa Inc | 1257 | 16,54724 | 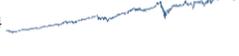 | 172 | 8 |
| DOWJONES_UNH | UnitedHealth Group Inc | 1257 | 16,62686 | 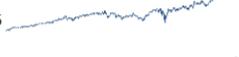 | 201 | 9 |
| DOWJONES_MCD | McDonald s Corp | 1257 | 16,73301 | 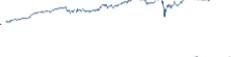 | 114 | 10 |
| DOWJONES_WMT | Walmart Inc | 1257 | 16,79936 | 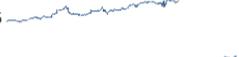 | 104 | 11 |
| DOWJONES_AXP | American Express Company | 1257 | 16,83917 | 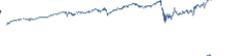 | 191 | 12 |
| DOWJONES_DIS | Walt Disney Co | 1257 | 16,85244 | 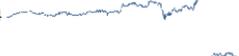 | 90 | 13 |
| DOWJONES_HON | Honeywell International Inc | 1257 | 16,86571 | 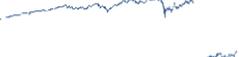 | 105 | 14 |
| DOWJONES_JPM | JPMorgan Chase & Co | 1257 | 16,86571 | 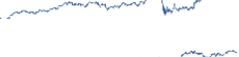 | 145 | 14 |
| DOWJONES_PG | Procter & Gamble Co | 1257 | 16,98514 | 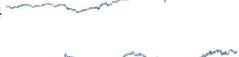 | 61 | 15 |
| DOWJONES_TRV | Travelers Companies Inc | 1257 | 17,1311 | 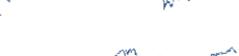 | 36 | 16 |
| DOWJONES_CSCO | Cisco Systems Inc | 1257 | 21,32431 | 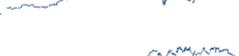 | 80 | 17 |
| DOWJONES_AMGN | Amgen Inc | 1257 | 21,33758 | 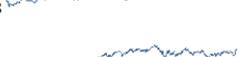 | 25 | 18 |
| DOWJONES_MRK | Merck & Co Inc | 1257 | 21,43047 | 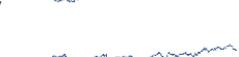 | 34 | 19 |
| DOWJONES_JNJ | Johnson & Johnson | 1257 | 21,47028 | 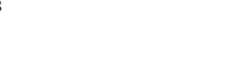 | 34 | 20 |

| Название индекса и код ценной бумаги (sticker) | Название компании (company) | Длина ряда исходных данных Цены закрытия (points) | Показатель сжатия ряда данных изменений Цены закрытия, % (variaty) | Микро-график ряда исходных данных Цены закрытия (microchart) | Рост Цены закрытия за период, % (growth) | Ранг аналитики (rank) |
|---|---|---|---|---|---|---|
| DOWJONES_INTC | Intel Corporation | 1257 | 21,70913 | 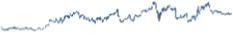 | 39 | 21 |
| DOWJONES_KO | Coca-Cola Co | 1257 | 21,7224 | 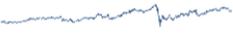 | 30 | 22 |
| DOWJONES_DOW | Dow Inc | 648 | 36,07796 | 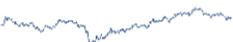 | 16 | 23 |
| DOWJONES_BA | Boeing Co | 1257 | {16,53397} | 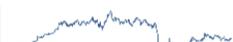 | 67 | 24 |
| DOWJONES_WBA | Walgreens Boots Alliance Inc | 1257 | {16,6534} | 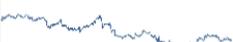 | -39 | 25 |
| DOWJONES_MMM | 3M Co | 1257 | {16,87898} | 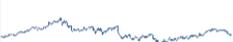 | 3 | 26 |
| DOWJONES_CVX | Chevron Corporation | 1257 | {21,31104} | 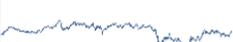 | 6 | 27 |
| DOWJONES_IBM | IBM Common Stock | 1257 | {21,32431} | 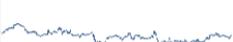 | -9 | 28 |
| DOWJONES_VZ | Verizon Communications Inc | 1257 | {21,80202} | 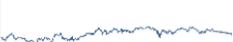 | 2 | 29 |

На этом ранжирование списка ценных бумаг по показателю волатильности доходности завершено. Рассчитанные указанным способом результаты ранжирования акций, входящих в фондовый индекс Dow Jones, представлены в первой из двух приведённых выше таблиц. В свою очередь результаты их ранжирования по показателю сжатия ряда данных изменений биржевой цены представлены во второй из двух приведённых выше таблиц.

При сравнении результатов ранжирования акций по волатильности доходности и сжатию изменений цены наблюдаются следующие особенности:

- значения ранга показателя волатильности не коррелируют со значениями ранга показателя сжатия;
- значения ранга показателя волатильности не коррелируют с показателем роста доходности цены за период (и предполагаемым наклоном тренда микрографика);
- увеличение ранга показателя сжатия (т.е. уменьшение величины сжатия) коррелирует с уменьшением показателя роста доходности цены за период (и уменьшением предполагаемого наклона тренда микрографика).

## 6      Заключение (Conclusion)

В широком понимании ранжирование ценных бумаг это процесс их упорядочивания по рыночным характеристикам в интересах инвесторов. К ранжированию как правило прибегают участники фондового рынка, оказывающие брокерские или консультационные услуги своим клиентам. Случается, что их методики ранжирования либо сложны, либо закрыты, либо недостаточно эффективны с точки зрения инвесторов. В данной работе представлена полностью и подробно описанная методика, которая сопровождается примерами расчётов показателей ранжирования на реальных биржевых данных. При принятии инвестиционных решений эту методику целесообразно использовать вкупе с другими инструментами технического, фундаментального, портфельного анализа ценных бумаг.

Предлагаемая методика как одно из средств обеспечения технического анализа предполагает возможность экстраполяции результатов ранжирования по завершённым периодам биржевой торговли на будущие периоды. Она ориентирована на поддержку долгосрочной (и в некоторой степени консервативной) стратегии вложения финансовых средств в рынок акций или ценных бумаг биржевых фондов. Инвестору, который пожелает воспользоваться этой методикой, следует синхронизировать по времени периодичность расчётов листингов ранжирования с периодичностью пересмотра своего инвестиционного портфеля.

Сформулированные выше высказывания о наблюдаемых в листингах ранжирования (которые представлены в качестве расчётных примеров) закономерностях целесообразно проверить численно (статистически) как на использованных в настоящей работе рядах данных, так и рядах данных большего количества акций, торгуемых на различных биржевых рынках, однако такая задача выходит за рамки описываемой здесь работы.

Данная статья может быть полезна для образовательных, исследовательских, консультационных целей, а также для целей разработки программного обеспечения и автоматизации трудоёмких расчётов листингов ранжирования инвестиционной привлекательности ценных бумаг по предлагаемому методу.

## 7      Ссылки (References)